\documentclass[prd,preprint,eqsecnum,nofootinbib,amsmath,amssymb,
               tightenlines,dvips]{revtex4}

\usepackage{color,latexsym,graphics,graphicx,epsfig} 
\usepackage{graphicx,graphics,color,latexsym}
\usepackage{bm}

\def\x{{\bf x}}

\def\p{{\bf p}}
\def\k{{\bf k}}
\def\q{{\bf q}}
\def\v{{\bf v}}

\def\u{{\bf u}}

\def\Im{{\rm Im}}

\def\x{{\bf x}}

\def\st{\begin{equation}}
\def\stp{\end{equation}}
\def\bg{\begin{eqnarray}}
\def\nd{\end{eqnarray}}
\def\Eq#1{Eq.~(\ref{#1})}
\def\app#1{Appendix~\ref{#1}}
\def\Fig#1{Fig.~\ref{#1}}
\def\Sect#1{Section~\ref{#1}}
\def\Ref#1{Ref.~\cite{#1}}

\def\llangle{\left\langle}
\def\rrangle{\right\rangle}

\def\tr{\operatorname{tr}}

\def\dpslash{\frac{d^3\p}{(2\pi)^3} \>}

\def\JJ{{\scriptscriptstyle JJ}}
\def\NN{{\scriptscriptstyle NN}}
\def\ppx#1{\frac{\partial}{\partial #1} }

\def\wb{\mathfrak{w}}
\def\qb{\mathfrak{q}}
\def\N{\mathcal{N}}
\def\Ell{\mathcal{L}}
\def\phiz{{\Phi}}
\def\phio{\phi_1}
\def\phiob{\overline{\phi_1}}

\def\g{{\bf g}}
\def\csq{c_s^2}
\def\gt{{\bf g}^{\scriptscriptstyle T}}
\def\gl{g^{\scriptscriptstyle L}}

\def\nott#1{\setbox0=\hbox{$#1$}                
   \dimen0=\wd0                                
   \setbox1=\hbox{/} \dimen1=\wd1               
   \ifdim\dimen0>\dimen1                        
      \rlap{\hbox to \dimen0{\hfil/\hfil}}      
      #1                                        
   \else                                        
      \rlap{\hbox to \dimen1{\hfil$#1$\hfil}}   
      /                                         
   \fi}                                         %

\advance\parskip 1.9pt
\advance\voffset -0.2in

\def\Eq#1{Eq.~(\ref{#1})}
\def\app#1{Appendix~\ref{#1}}
\def\Fig#1{Fig.~\ref{#1}}
\def\Sect#1{Section~\ref{#1}}
\def\Ref#1{Ref.~\cite{#1}}

\begin{document}

\title{Finite Temperature Spectral Densities of Momentum and
R-Charge Correlators in $\N=4$  Yang Mills Theory}

\author{Derek Teaney}
\affiliation
    {%
    Department of Physics \& Astronomy,
    SUNY at Stony Brook,
    Stony Brook, New York 11764, USA
    }%

\date{\today}

\begin{abstract}
We compute  spectral densities of momentum and R-charge correlators in thermal
$\N=4$ Yang Mills at strong coupling using the AdS/CFT correspondence.  For
$\omega \sim T$ and smaller, the spectral density differs markedly from
perturbation theory; there is no kinetic theory peak. For large $\omega$, the
spectral density oscillates around the zero-temperature result with an
exponentially decreasing amplitude. Contrast this with QCD where  the spectral
density of the current-current correlator approaches the zero temperature
result like $(T/\omega)^4$.  Despite these marked differences with perturbation
theory, in Euclidean space-time the correlators differ by only $\sim 10\%$ from
the free result. The implications for Lattice QCD measurements of transport are
discussed.
\end{abstract}

\maketitle


\section{Introduction}

The experimental relativistic heavy ion program has produced a variety
of evidences which suggest that a Quark Gluon Plasma (QGP) has been
formed at the Relativistic Heavy Ion Collider (RHIC)
\cite{Bellwied:2005kq,Adcox:2004mh}.  The relative 
success of hydrodynamic approaches 
\cite{Ollitrault:1992bk,Hirano:2004er,Teaney:2001av,Kolb:2000fh,Huovinen:2001cy}
suggests that the mean free path of the QGP is
of order the thermal wave length $\sim T$ \cite{Molnar:2001ux}. This bold
inference requires further theoretical and experimental corroboration.

Theoretically,  transport coefficients have
been computed in the perturbative QGP using 
kinetic theory \cite{Baymetal,AMY6}. However, it is important to
obtain non-perturbative estimates 
for the transport properties  of the medium.
Kubo formulas relate hydrodynamic 
transport coefficients  to the small frequency behavior
of correlation functions of conserved 
currents in real time \cite{Forster,BooneYip}.
In imaginary time correlation functions may be obtained from
Lattice QCD measurements.
Therefore, with a  good model for  the spectral density,
one could hope to obtain a reasonable non-perturbative 
estimate for the transport
properties of the QGP from Lattice QCD measurements.
Recently, attempts to extract the shear
viscosity \cite{nakamura97,nakamura05} and electric conductivity \cite{gupta03} have been made.  More generally, the current-current 
correlator is being studied actively in order
to extract interesting physical effects, {\it e.g.}  dilepton emission, Landau damping, heavy quark diffusion, and in medium modifications of $J/\psi$.  
\cite{Peter_light,Peter_heavy,teaneyd,Aarts:2005hg,Mocsy:2005qw,Nakahara:1999vy}.

Whenever there is a large separation between the transport
time scales and the temperature $T$, Euclidean correlators
are remarkably insensitive 
to transport coefficients \cite{aarts02,teaneyd}. This 
has been studied in perturbation theory \cite{aarts02}, 
where the transport time scale is $1/(g^4 T)$, and in a 
theory with a heavy quark, where the  diffusion time scale 
is $M/T^2$ or longer \cite{teaneyd}. 
This insensitivity  to transport is because the dominant transport contribution to the Euclidean
correlator is an integral over the spectral density which
is governed by a sum rule -- the $f$-sum rule. Whenever,
there is a quasi-particle description, the leading transport contribution
to the Euclidean current-current correlator is proportional
to the static susceptibility times an average thermal
velocity squared of the quasi-particle. 
Only with very accurate measurements and a solid understanding of the
spectral density can more transport information be gleaned from
lattice correlators.
 
When there is no separation between the inverse temperature
and transport time scales this conclusion requires further study. 
$\N=4$ Super Yang Mills (SYM) theory with a large number 
of colors and large 't Hooft coupling provides
a theory where there is no scale separation and where  the
spectral density can be computed if the Maldacena conjecture
is accepted \cite{Maldacena:1997re}. This conjecture states that  
 $\N=4$ SYM is dual to classical type IIB supergravity
on an $AdS\times S_{5}$ background 
\cite{Maldacena:1997re,Gubser:1998bc,Witten:1998qj}. 
(See \Ref{Aharony:1999ti} for a review.)
By solving the classical supergravity equations motion we may
deduce the retarded correlators of strongly coupled $\N=4$ 
SYM. 

Using the correspondence, Son, Policastro, and Starinets
deduced the shear viscosity of strongly coupled $\N=4$ SYM
\cite{Policastro:2001yc}.
Indeed the momentum diffusion constant is
remarkably small, $\eta/(e + p) = 1/4\pi T$. 
This result and subsequent calculations  have led to the conjecture that the shear viscosity to
entropy ratio is bounded from below, $\eta/s \ge 1/4\pi$ 
 \cite{Kovtun:2004de,Buchel:2004qq,Policastro:2001yc}.
The correspondence was also used to calculate the R-charge 
diffusion coefficient, $D=1/(2\pi T)$ \cite{Policastro:2002se}.  
This transport coefficient is also remarkably small
although there is no conjectured bound associated with
this medium property.
It is interesting to compute the full spectral densities
of these correlators following the framework set up in these works.  
With the spectral density in hand, the feasibility of
extracting transport from the Lattice in a strong coupling
regime can be assessed.

\section{Stress Tensor  and  R-charge Correlations}

\subsection{Stress Tensor Correlations}
\label{stress-tensor}

First let us recall the definitions \cite{Forster}. 
The retarded correlators
of $T^{00}$ and $T^{0i}$  are defined as follows
\bg
\chi_{ee} (\k,\omega) &\equiv& \int_{-\infty}^{\infty} dt\, \int d^3\x\, e^{+i\omega t - i\k\cdot \x } \, 
i\theta(t) \llangle \left[ T^{00}(\x,t), T^{00}(0,0) \right] \rrangle \, , \\
\chi_{gg}^{ij}(\k, \omega) &\equiv& \int_{-\infty}^{\infty} dt \int d^3\x\, e^{+i\omega t- i\k \cdot \x }\,
i\theta(t) \llangle \left[ T^{0i}(\x,t), T^{0j}(0,0) \right] \rrangle \, .
\nd
$\chi_{gg}^{ij}$ is then decomposed into longitudinal 
and transverse pieces
\st
  \chi_{gg}^{ij}(\k,\omega) = \frac{k^{i}k^{j}}{k^2}\,\chi_{gg}^{L}(\k,\omega)  
   + \left(\delta^{ij} - \frac{ k^i k^j}{k^2}\right) \, \chi_{gg}^T(\k,\omega) \, .
\label{chigg}
\stp
Through linear response, hydrodynamics makes a definite prediction for
these correlators at small $k$ and $\omega$ (see \app{modes} for a
review)
\bg
   \chi_{gg}^L (\k, \omega) &=& (e+ p)\,\frac{i\omega \Gamma_s k^2 - (c_s k)^2} 
{\omega^2 - (c_s k)^2 + i\omega \Gamma_s k^2}  \, ,\\
   \chi_{gg}^T (\k, \omega) &=& \frac{ \eta k^2}{-i\omega + \frac{\eta k^2}{e+p} } \, ,
\label{chigg_hydro}
\nd 
where $e$ is the energy density, $p$ is the pressure, $\eta$ is 
the shear viscosity, $\zeta$ is the bulk viscosity, and  $\Gamma_s = \frac{\frac{4}{3}\eta + \zeta}{e + p}$ is the sound attenuation length.
Using energy conservation we deduce
\st
   \chi_{ee}(\k, \omega) = \frac{\omega^2}{k^2} \chi_{gg}^{L} \, .
\stp
Similarly we define the correlators of momentum fluxes
\st
    \chi_{\tau\tau}^{iljm}(\k,\omega) = \int_{-\infty}^{\infty} dt \,\int d^3\x\, e^{+i\omega t - i\k\cdot\x}\, i\theta(t)\llangle [ T^{il}(\x,t), T^{jm}(0,0) ] \rrangle  \, .
\stp
Then using momentum conservation, we deduce that
\bg
\label{current_conserve_1}
  \chi^{xxxx}_{\tau\tau} &\equiv& \frac{k^i k^l k^j k^m}{k^4} \chi_{\tau\tau}^{il,jm}(\k,\omega) = \frac{\omega^2}{k^2}\, \chi_{gg}^L(\k,\omega)  \, , \\
\label{current_conserve_2}
  \chi^{yxyx}_{\tau\tau} &\equiv& \frac{1}{2} \left(\delta^{ij} -\frac{k^i k^j}{k^2} \right) \,\frac{k^l k^m}{ k^2}\, \chi_{\tau\tau}^{il,jm}(\k,\omega) = \frac{\omega^2}{k^2}\, \chi_{gg}^T(\k,\omega)  \, . 
\nd
The notation follows by assuming that $\k$ points in the $x$ direction.

We will concentrate on the transverse piece of the 
momentum correlator. We define the spectral density    
as follows
\st
   \rho_{\tau\tau}^{yxyx}(\k,\omega) = \frac{\mbox{Im} \chi_{\tau\tau}^{yxyx}(\k,\omega)}{\pi}   \, .
\stp
Using the functional form in \Eq{chigg_hydro} we deduce the Kubo 
formula
\st
    \lim_{\omega\rightarrow0} \lim_{k\rightarrow0} \frac{\pi \rho_{\tau\tau}^{yx,yx}(\k,\omega) }{\omega} = \eta \, .
\stp
Euclidean correlators $G^{yxyx}_{\tau\tau}(\x,\tau) =\llangle T^{yx}_{E}(\x,\tau) T^{yx}_{E}(0) \rrangle$ can be deduced from the spectral density
\st
  G_{\tau\tau}^{yxyx}(\k,\tau) =  \int_0^{\infty} d\omega \, \rho_{\tau\tau}^{yxyx}(\k,\omega) \frac{ \cosh\left( \omega\left(\tau - \beta/2\right)\right) }
{\sinh\left(\omega \beta/2\right) } \, .
\label{spectral_rep}
\stp
Further explanation of the Euclidean definitions and conventions
adopted in this work may be found in \cite{teaneyd}.

The retarded correlator $\chi_{\tau\tau}^{yxyx}(\k,\omega)$  
may computed using the AdS/CFT correspondence.   The stress tensor
couples to metric fluctuations $h_{xy}$ in the background of
the black hole. 
We start with equation Eq.~(6.6) of Policastro, Son and Starinets \cite{Policastro:2002se} for the fluctuation $\phi\equiv h^x_y$  in the 
gravitational background. The linearized equations of 
motion for this field in the background are
\begin{eqnarray}
\label{stareq}
   \phi''_k - \frac{1 + u^2}{u f} \phi'_k  +  
\frac{\wb^2 - \qb^2 (1-u^2) }{u f^2} \phi_k= 0 \, .
\end{eqnarray}
Here $\phi_k$ is the Fourier transform of $\phi$ with
respect to four momentum $k = (\omega,0, 0, q)$ and
we have defined 
 $\wb \equiv \omega/(2\pi T)$, 
$\qb \equiv q/(2\pi T)$, and
$f \equiv (1-u^2)$. $u=1$ corresponds to the  
the event horizon of the black hole; $u=0$ corresponds to
spatial infinity.
We will integrate this equation and
set $\qb=0$ in what follows.

Let us indicate the general strategy. 
Near the horizon $u=1$, we develop the series solutions \cite{Morse}
\bg
   \phio &=& (1-u)^{-i\wb/2} \left[
1 + (1-u) \left( \,\frac{\frac{3\wb^2}{4} - i \frac{\wb}{4} + i \frac{\wb^3}{2} }{1 + \wb^2}\right) + \dots \right ]\, , \\ 
   \phi_2 &=& \phiob \, , 
\nd 
where ${\phiob}$ denotes the complex conjugate of $\phio$. 
The physical solution is $\phio$ and not $\phiob$.
Similarly near spatial infinity $u=0$, we can develop a power series solution of this equation.
These solutions are
\bg
\phiz_1 &=& u^2 \left[ 1 - \frac{\wb^2}{3} u + 
\dots \right] \, , \\
\phiz_2 &=& -\frac{\wb^4}{2}\,\log(u) \phiz_1 + \left[ 1 + \wb^2u 
 +  \left(\frac{1}{2} - \frac{29}{72}\wb^4 \right) u^2 + 
\dots \right] \, .
\nd
We note that $\phiz_1$ and $\phiz_2$ are real functions.
Generally $\phiz_i$ will be a linear combination $\phio$ and 
$\phiob$. 
Following \cite{Policastro:2002se},
the physically correct solution of \Eq{stareq} is 
$\phio$ which is a linear combination $\phiz_1$ and $\phiz_2$:
\st
\label{linearco}
     \phio =  A \phiz_2  + B \phiz_1 \,.
\stp
We may set $A=1$ which is equivalent to the often assumed
normalization condition $\left.\phi_1(u)\right|_{u=0} = 1$\,.
Then the retarded Greens function is related by the
correspondence to $\phio$
\st
    \chi_{\tau\tau}^{yxyx}(\q, \omega)  =  \frac{\pi^2 N^2 T^4}{4} \;  
\lim_{u\rightarrow0} \frac{f}{u}(\phi_{-k})_1 \partial_u (\phi_k)_1  \, ,
\label{GRTxy}
\stp
where $(\phi_k)_1$ denotes the $\phi_1$ which
solves \Eq{stareq} with $k=(\omega, \q)$.
Using the  power series expansion near $u\approx0$, 
and the fact that $\phiz_1$ and $\phiz_2$ are real, 
we have the spectral density\footnote {
As is common in the lattice community the spectral density is defined so that 
$\left. \frac{\pi \rho(\omega)}{\omega} \right|_{\omega=0}= \eta$. 
The retarded greens function differs by an overall sign from \Ref{Policastro:2002se} in order to 
conform with \Ref{teaneyd}. }
\bg 
    \frac{\pi \rho_{\tau\tau}^{yxyx}(\omega)}{\omega} \equiv \frac{\Im \chi_{\tau\tau}^{yxyx}(\omega,\q)}{\omega} &=& \frac{ \pi^2 N^2 T^4}{4} \,\Im B\, \lim_{u\rightarrow0} \frac{f}{u} \phiz_2 \partial_u (\phiz_1) \, , \\
\label{spectralD}
                         &=& \frac{\pi^2 N^2 T^4}{4} \, 2\,\Im B \, . 
\nd
Thus the algorithm may divided into four steps:
(1) Start at small $u \approx 0$ and use the power series to
find the initial condition and first derivative of 
$\phiz_1$ and $\phiz_2$. 
(2) Integrate $\phiz_1$ and $\phiz_2$ forward to determine
the solution close to $u\approx1$.
(3) Using the power series close to $u\approx1$  determine
the  linear combination of $\phio$ and $\phiob$ that
is $\phiz_1$. Also do this for $\phiz_2$. 
(4) Then determine
the linear combination of $\phiz_1$ and $\phiz_2$ that
is $\phio$. This fixes $B$ in Eqs.~(\ref{linearco}) and (\ref{spectralD}), and consequently determines the spectral density at the frequency
specified in \Eq{stareq}.

This algorithm was followed and the resulting spectral density
is shown in \Fig{rho}(a).  
When the frequency $\wb$ is large we may follow this 
algorithm analytically and obtain the zero temperature result
at large frequency.  This 
is done in \app{large_omega} and the resulting solution
provides a good check of the numerical work. Further discussion
of this figure in provided in  \Sect{summary_sect}.
 
\subsection{R-Charge Diffusion}

A similar program may be followed to calculate the 
spectral density of the current-current correlator.
First let us recall the definitions.  
Through linear response \cite{Forster, teaneyd},
the diffusion equation
predicts that the density-density correlator
\st
      \chi_\NN(\k, \omega) = \int_{-\infty}^{\infty} dt\,\int d^3\x\, e^{+i\omega t - i\k\cdot \x} \,  
   i\theta(t)\llangle \left[J^0_a(\x,t) ,J^0_a(0,0) \right]\rrangle \qquad
\mbox{(no $a$ sum)} \, , 
\stp
has the following form at small $k$ and $\omega$
\st
     \chi_\NN (k,\omega)  = \frac{\chi_s Dk^2}{-i\omega + D\,k^2} \, ,
\label{diffusion}
\stp
where $\chi_s$ is the static R-charge susceptibility.
The spatial current-current correlator is defined  similarly
\st
    \chi_\JJ^{ij}(\k, \omega) = \int_{-\infty}^{\infty} dt\,\int e^{+i\omega t - i\k\cdot \x}  
   \,i\theta(t) \,\llangle \left[J^i_a(\x,t), J^j_a(0,0)\right] \rrangle \qquad 
\mbox{(no $a$ sum)} \, ,
\stp
and may be broken up into longitudinal and transverse 
components
\st
\label{chiL}
   \chi_\JJ^{ij}(\k,\omega) = \left(\frac{k^i k^j}{k^2} - \delta^{ij}\right)\, \chi_\JJ^T(\k,\omega)
   + \frac{k^i\,k^j}{k^2}\, \chi_\JJ^L(\k,\omega) \, .
\stp 
The density-density correlator can be related to the 
longitudinal  current-current correlator 
\st
\label{chiJJ}
\frac{\omega^2}{k^2} \chi_{\scriptscriptstyle NN}(\k,\omega) = \frac{k^ik^j}{k^2} \chi_\JJ^{ij}(\k,\omega) = \chi_\JJ^L(\k,\omega)\,.
\stp
For $\k=0$ there is no distinction between the longitudinal and transverse parts
and therefore  for $k \ll T$,  $\chi_\JJ^L(\k,\omega) \simeq
\chi_\JJ^T(\k,\omega)$. 

The computation of these correlators at small $\k$ and $\omega$
has already been performed by Policastro, Son, and Starinets \cite{Policastro:2002se}. 
From their computation (see Eq.~5.17b of that work) 
and the functional form of \Eq{diffusion}, the
diffusion coefficient and static susceptibility are
\st
   D = \frac{1}{2\pi T}\,, \qquad \mbox{and} \qquad \chi_s = \frac{N^2 T^2}{8} \, .
\stp

To extend this computation to finite frequency  
it is simplest to compute $\chi_\JJ^T(0, \omega)$ which
is uncoupled from the other modes.  Since $\k = 0$,  
$\chi_{JJ}^{T}(0,\omega)$  is equal to  $\chi_\JJ^L(0, \omega)$.
Perturbations in the R-charge current couple to the 
Maxwell field and the equations of motion 
for the Maxwell field  in the gravitational background 
have been worked out. Consider the Maxwell field $\phi$ transverse
to $\q$, {\it i.e.} if $\q$ points in the $z$ direction then   
$\phi = A_{x}$. The equation of motion for the 
Fourier components  of $\phi$ in
the gravitational background reads 
\st
  \phi_k'' + \frac{f'}{f}\,\phi_k' + \frac{1}{uf} 
\left( \frac{\wb^2}{f} - \qb^2 \right) \phi_k = 0 \;.
\stp
with $f = 1-u^2$ and as before $k=(\omega, 0,0,q)$, $\wb = \omega/(2\pi T)$ and $\qb = q/(2\pi T)$. 
In what follows we set $\qb=0$.

The procedure mirrors the stress tensor computation.  
Near $u=1$ we determine a power series solution
\bg
   \phio &=& (1-u)^{-i\frac{\wb}{2}}\left[1 + \dots\right]\, , \\
   \phi_2 &=& \phiob \, .
\nd
$\phio$  is the physical solution and not $\phiob$.
Near $u=0$ we determine the power series 
\bg
    \phiz_1 &=& \wb u + \dots \, , \\
    \phiz_2 &=& 1 +  \dots \; .
\nd
Integrating from $u\approx0$ to $u\approx1$ we 
determine the linear combination of $\phiz_1$ and 
$\phiz_2$ that is $\phio$
\st
    \phio = \phiz_2 + B\,\phiz_1 \, .
\stp
Then the correspondence states that
\st
    \chi_\JJ^T(0, \omega) =  \frac{N^2 T^2}{8}\lim_{u\rightarrow 0} f (\phi_{-k})_1\partial_u 
     (\phi_k)_1 \; .
\stp
Since $\phiz_2$ and $\phiz_1$ are real 
and since $\phiz_2$ starts with one, the spectral density is
\st
    \frac{\pi \rho_\JJ^{T,L}(0,\omega)}{\omega} 
\equiv \frac{\mbox{Im} \chi_{JJ}^{T}(0,\omega)}{\omega} =  
  \frac{N^2 T}{16 \pi}\, \mbox{Im}\,B \, .
\stp

\section{Results}
\label{summary_sect}
\begin{figure}[t]
\begin{center}
\includegraphics[height=3.0in,width=3.2in]{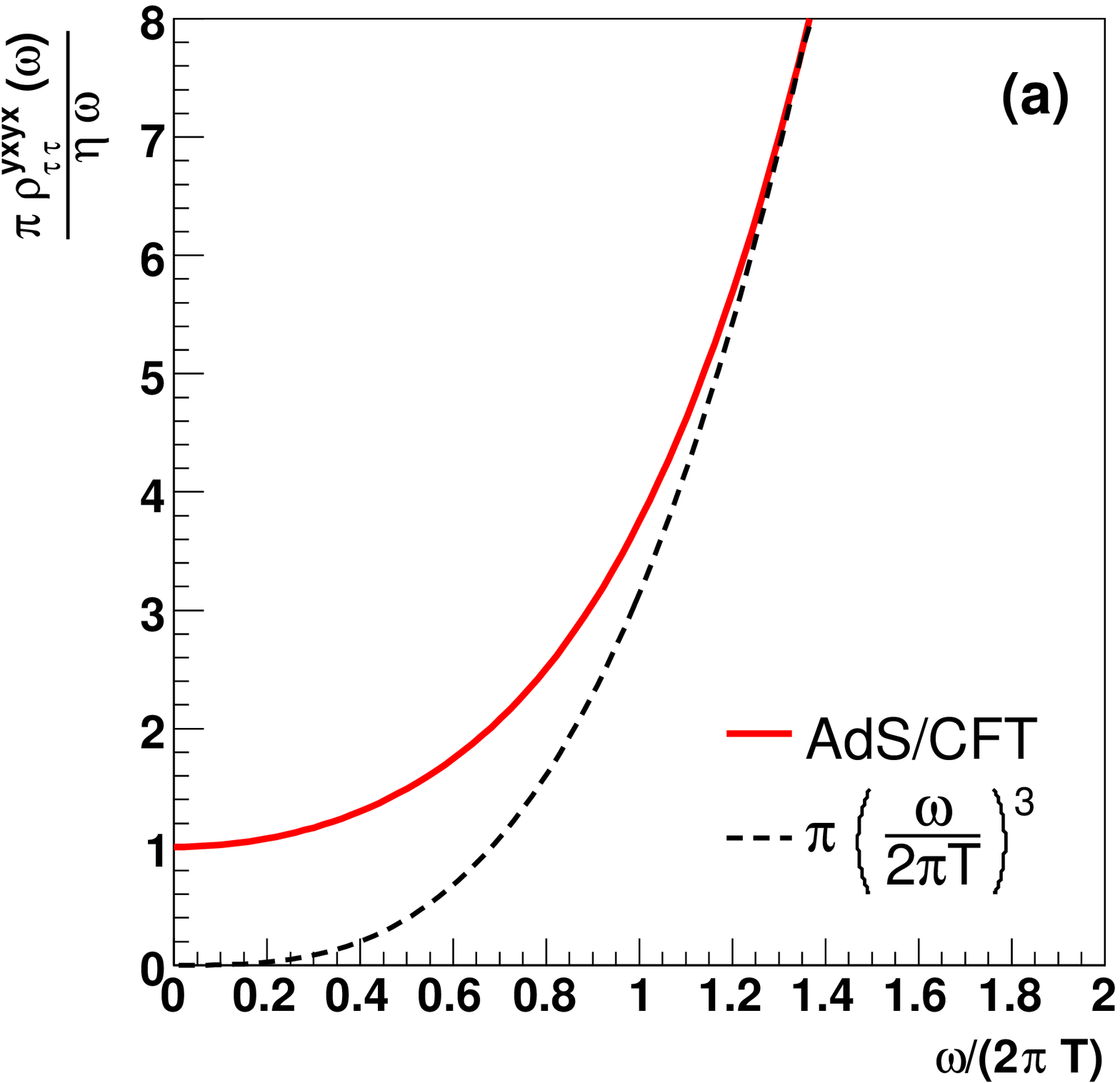}
\hfill
\includegraphics[height=3.0in,width=3.2in]{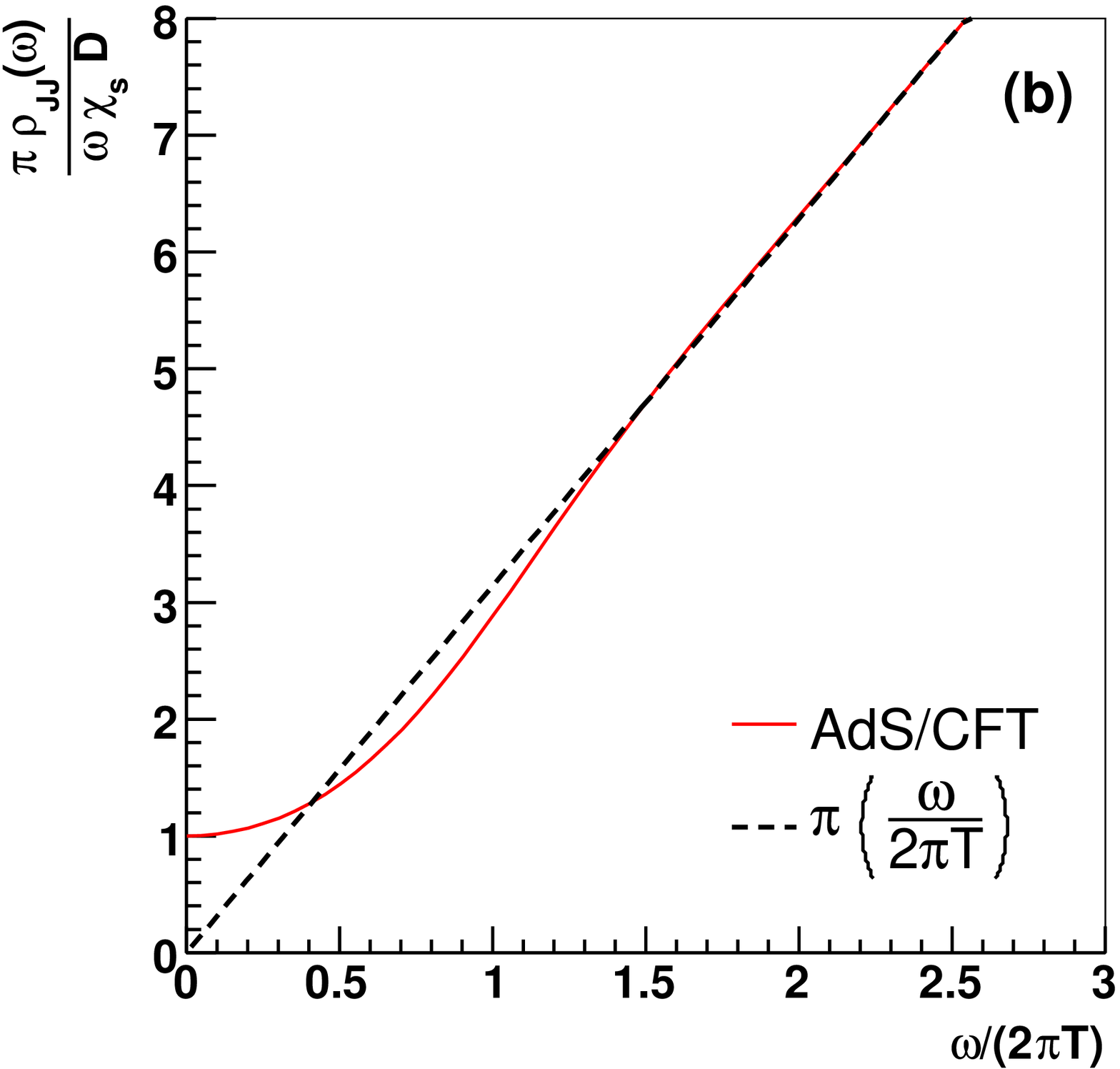}
\caption{
(a) The spectral density of the stress energy tensor,
$\pi \rho_{\tau\tau}^{yxyx}(\omega)/\omega$ 
normalized by the shear viscosity, 
$\eta_{\scriptscriptstyle AdS} = \pi N^2 T^3/8$. 
(b) The spectral density of the current-current correlator,
$\pi \rho_{JJ}/\omega$ normalized by  $\chi_s D = N^2 T^2/16\pi T $.
In both cases the dashed curves show the 
zero temperature results (\Eq{rho_tau_0T} and \Eq{rho_j_0T}) normalized 
by the same factors. Due to a non-renormalization theorem in these channels, 
the zero temperature spectral densities in the free and interacting theories are
equal \cite{Freedman:1998tz,Chalmers:1998xr}. 
At finite temperature the kinetic theory peak does not exist in the strongly interacting theory.
}
\label{rho}
\end{center}
\end{figure}
\begin{figure}[t]
\begin{center}
\includegraphics[height=3.0in,width=3.2in]{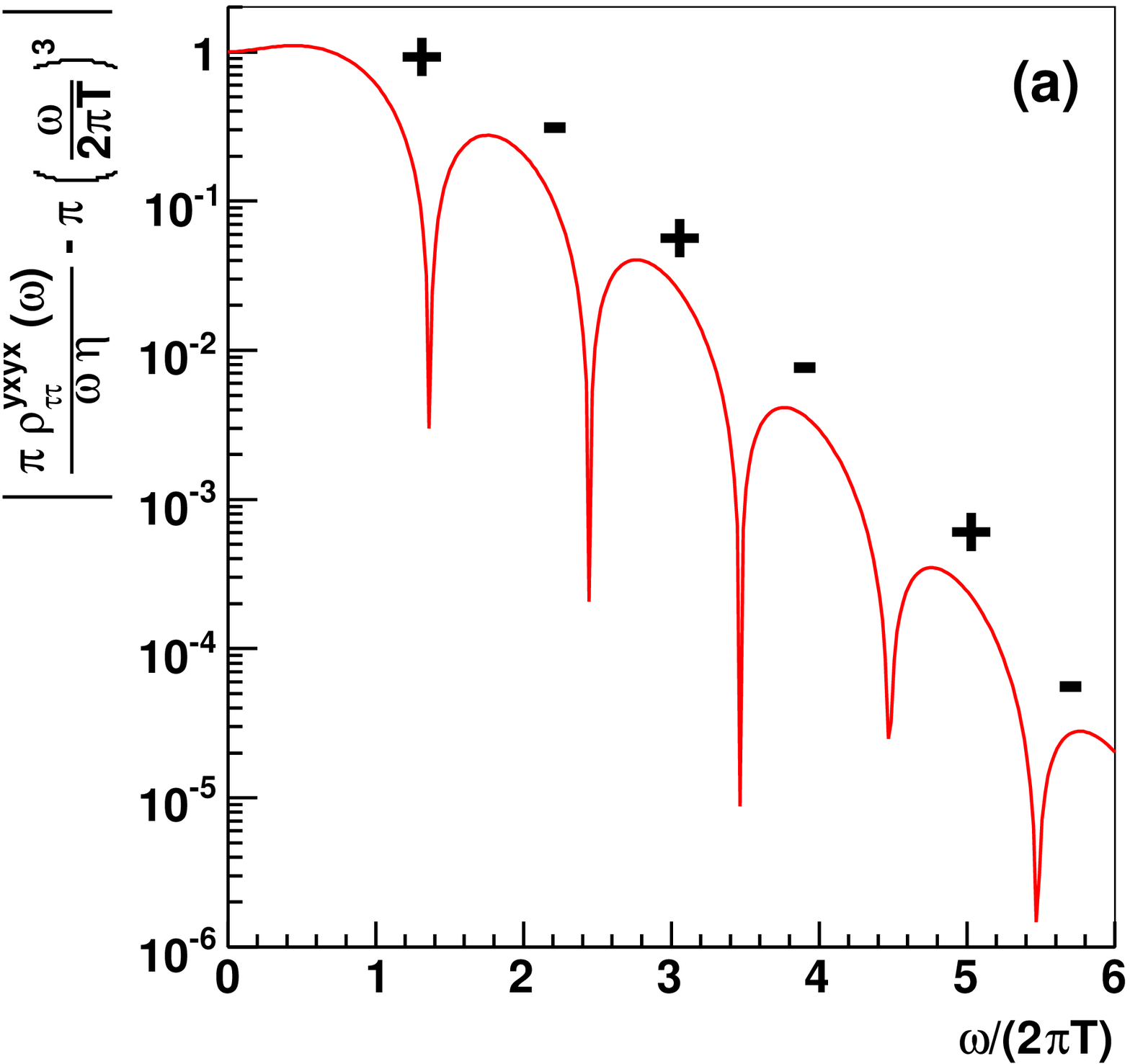}
\hfill
\includegraphics[height=3.0in,width=3.2in]{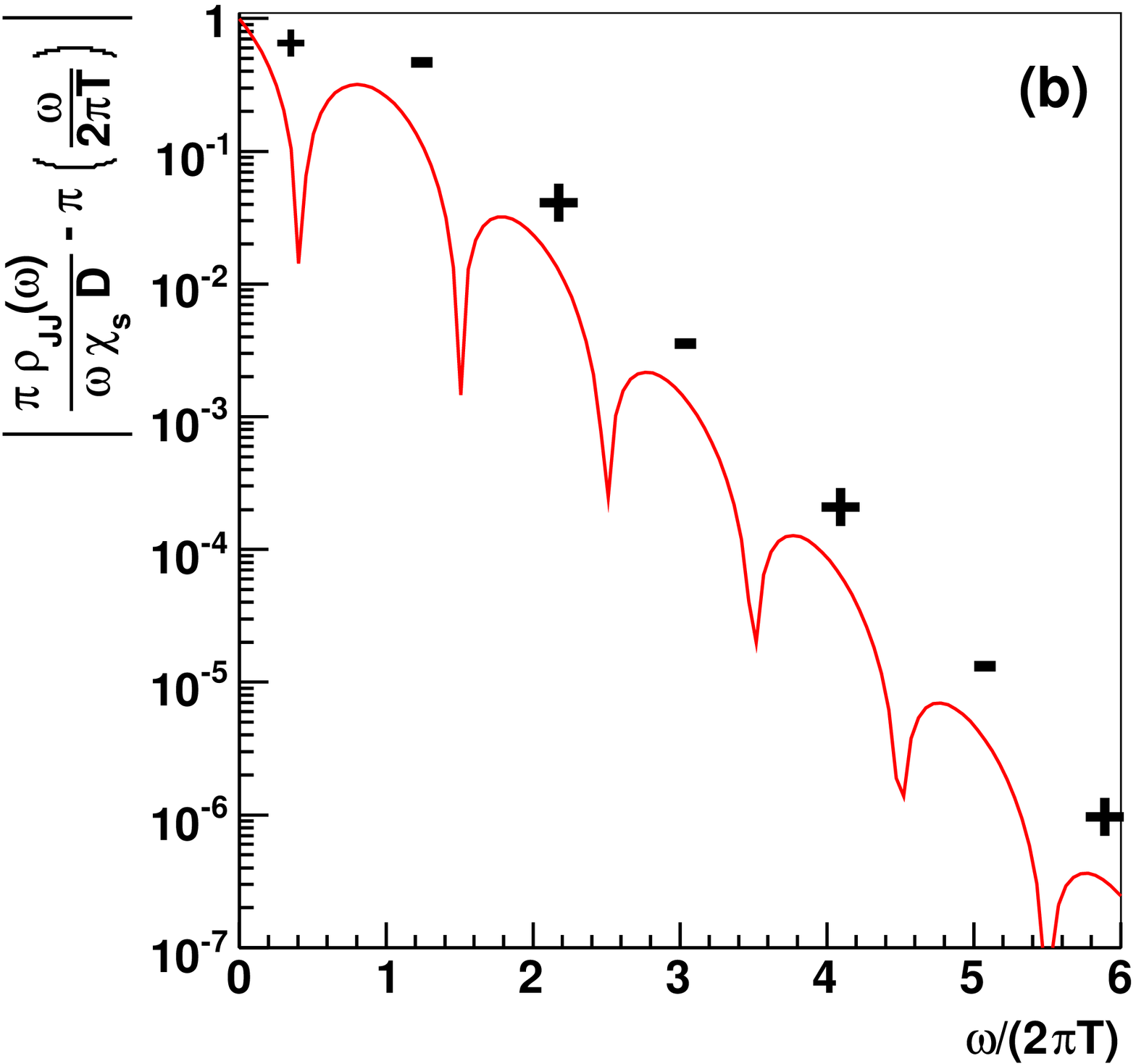}
\caption{
The spectral density of the (a) stress energy tensor and
(b) current-current correlators as in \Fig{rho} 
but 
the zero temperature result has been subtracted
and the absolute value taken. The plus or minus indicates
the sign.  The finite temperature spectral densities 
oscillate around the zero temperature result with
exponentially decreasing amplitude. 
}
\label{rho-oscillate}
\end{center}
\end{figure}
\begin{figure}[t]
\begin{center}
\includegraphics[height=3.0in,width=3.2in]{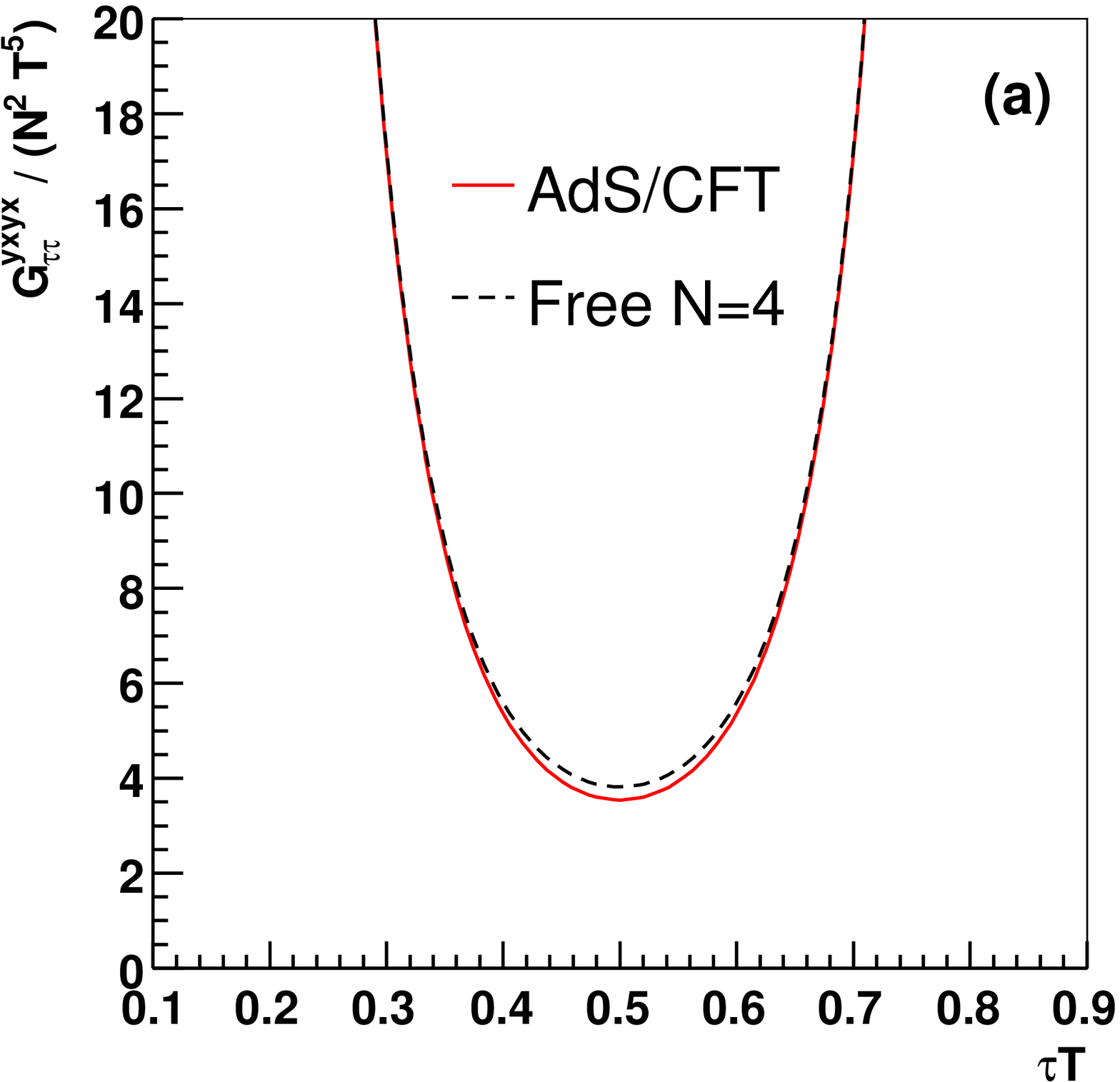}
\hfill
\includegraphics[height=3.0in,width=3.2in]{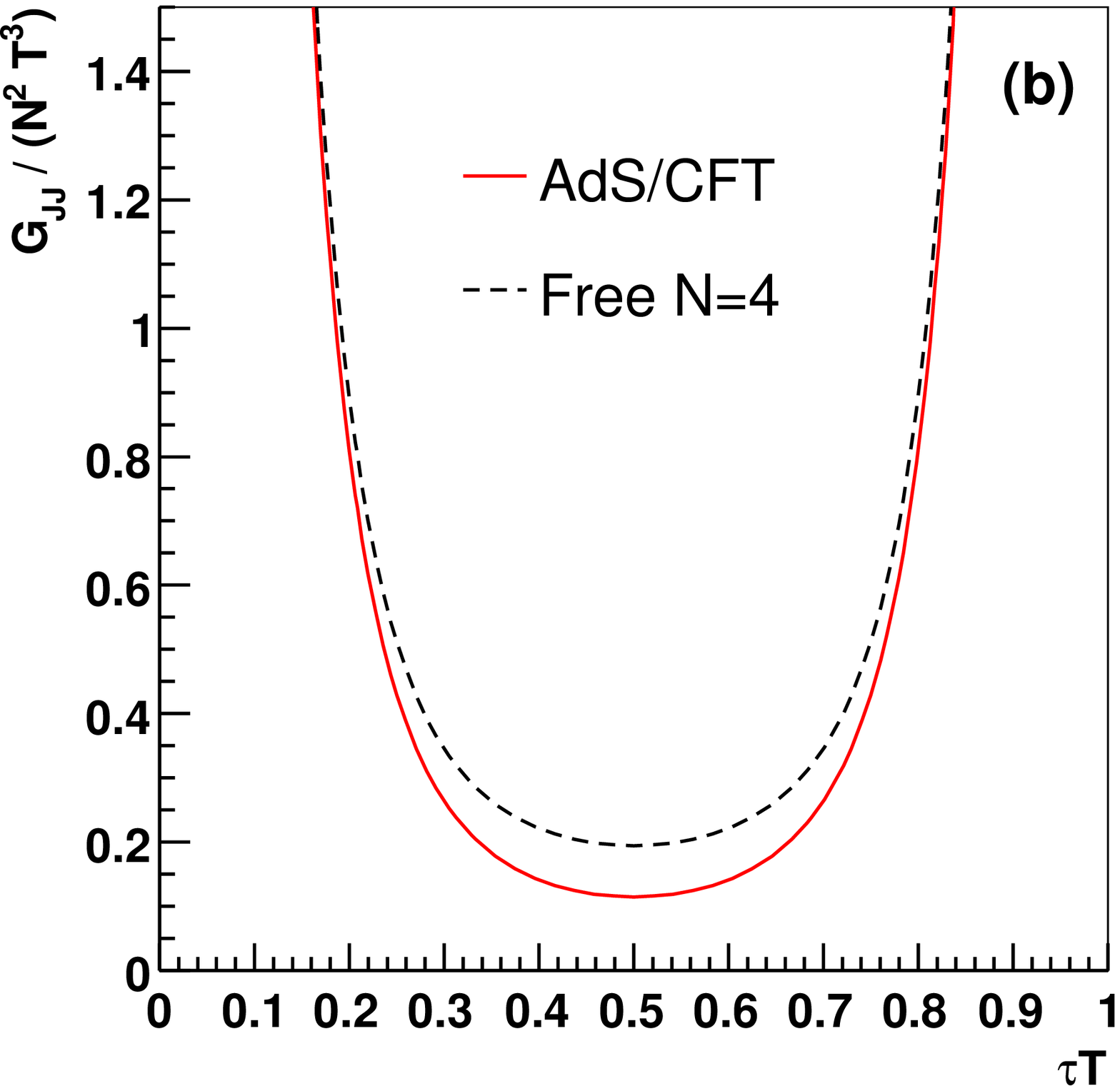}
\caption{
The Euclidean correlator for the (a) stress energy 
tensor correlator, $G^{yxyx}_{\tau\tau}$, and the 
(b) current-current correlator, $G_{JJ}$. The dashed curves 
show the free result for these Euclidean correlators.
}
\label{euclid}
\end{center}
\end{figure}

The spectral density of the stress energy tensor
\st
   \rho^{yxyx}_{\tau\tau}(\omega) = \frac{1}{2\pi} \int_{-\infty}^{\infty} dt\, e^{+i\omega t} \int d^3\x\, \llangle \left[ T^{yx}(\x,t), T^{yx}({\bf 0},0) \right] \rrangle \, ,
\stp
is shown in \Fig{rho}(a). Similarly
the spectral density for the R-charge current-current correlator
\st
   \rho_{\JJ}(\omega) = \frac{1}{2\pi} \int_{-\infty}^{\infty} dt\, e^{+i\omega t} \int d^3\x\, \llangle \left[ J^{x}_{a}(\x,t), J^{x}_{a}({\bf 0},0) \right] \rrangle   \qquad  \mbox{(no $a$ sum)} \, ,
\stp
is shown in \Fig{rho}(b). These are normalized so that
\st
  \left. \frac{\pi \rho_{\tau\tau}^{yxyx}(\omega)}{\omega} \right|_{\omega = 0} = \eta\, , \qquad  \mbox{and} \qquad
  \left. \frac{\pi \rho_{\JJ}(\omega)}{\omega} \right|_{\omega = 0} = \chi_s D,
\stp
where $\eta$ is the shear viscosity, $\chi_s$ is the 
static R-charge susceptibility, and $D$ is the R-charge
diffusion coefficient.

The remarkable feature of these spectral functions 
is the absence of any distinction between the 
 transport time scales and the continuum time 
scales. 
For comparison, consider the  
spectral density of the stress energy 
tensor in  the free theory as worked out in \app{free_section}. The full spectral density  
is a sum of the gauge, fermion, and 
scalar contributions
\begin{eqnarray}
\left(\rho_{\tau\tau}^{yxyx}\right)_{\rm gauge}
&=& 
\left(\frac{e + p}{5}\right)_{\scriptscriptstyle A} \omega \delta(\omega) +  
\frac{N^2}{160\pi^2} \,\left(1 + 2\,n_B\left(\omega \beta/2 \right)\right) \, \omega^4  \,,
\label{freerhotautaua_text} \\
\left(\rho_{\tau\tau}^{yxyx}\right)_{\rm fermion} &=&   
\left(\frac{e + p}{5}\right)_{\scriptscriptstyle \lambda} \omega \delta(\omega) +  
\frac{N^2}{160\pi^2} \,\left(1 - 2\,n_F\left(\omega \beta/2 \right)\right) \, \omega^4  \,, \\
\left(\rho_{\tau\tau}^{yxyx}\right)_{\rm scalar}  &=&   
\left(\frac{e + p}{5}\right)_{\scriptscriptstyle X} \omega \delta(\omega) + 
\frac{N^2}{320\pi^2} \,\left(1 + 2\,n_B\left(\omega \beta/2 \right)\right) \, \omega^4  \,,
\label{freerhotautauc_text}
\end{eqnarray}
where for example $(e+p)_{\scriptscriptstyle X}$ is the partial
enthalpy due to the free scalar fields -- see \app{free_section}. 
The free spectral function for the current-current 
correlator has a similar structure and is recorded in the
appendix.
Notice the delta function at the origin. 
According to the $f$ sum rule (see below),
perturbations will smear  the delta-function at the origin
but will not change the integral under the peak. 
The peak indicates a large separation between 
the transport and temperature scales and is
what makes kinetic theory possible.
It is hopeless to apply a quasi-particle description
to the strongly coupled $\N=4$ case.

Also shown in \Fig{rho}(a) and (b) is the spectral density at
zero temperature for the stress tensor and R-charge correlators.
At zero temperature, the strongly coupled spectral densities 
are {\it identical} to the free
$\N=4$ spectral densities as required by a non-renormalization
theorem \cite{Freedman:1998tz,Chalmers:1998xr}.  \Fig{rho-oscillate}(a) and (b) 
show the absolute value of the difference
between the finite temperature and zero temperature results
on a log plot. The figure shows that the finite temperature result oscillates around
the zero temperature result with exponentially decreasing amplitude.
This is quite unusual and the 
exponential decrease is not due to any obvious Boltzmann factor. 
In QCD for instance, the spectral 
density of the current-current correlator approaches the 
zero temperature correlator as\footnote{ There appears to be a discrepancy between 
two independent calculations; \Ref{Majumder:2001iy} finds that
the difference falls as $\left(T/\omega\right)^2$.
},\, 
$\left(T/\omega\right)^4$ \,
\cite{Altherr:1989jc,Baier:1988xv}. 
The author has no explanation for the strong
coupling results in $\N=4$ SYM.

Finally, we may determine the corresponding Euclidean 
correlators by integrating the spectral density. For 
instance $G^{yxyx}_{\tau\tau}$ is determined from $\rho^{yxyx}_{\tau\tau}(\omega)$
\bg
\label{gyxyx}
    G^{yxyx}_{\tau\tau}(\tau) &=& \int d^3\x  
  \llangle T^{yx}_E(\x,\tau) T^{yx}_E({\bf 0}, 0) \rrangle \, ,\\
                  &=&                       
   \label{gyxyx_2}
   \int_0^{\infty} d\omega \, \rho_{\tau\tau}^{yxyx}(\omega) \frac{ \cosh\left( \omega\left(\tau - \beta/2\right)\right) }
{\sinh\left(\omega \beta/2\right) } \, .
\nd
The resulting Euclidean correlators for the 
stress tensor and current-current correlators are shown
in \Fig{euclid}(a) and (b) respectively. 
In each figure the strongly coupled case is compared
to the free theory worked out in \app{free_section}.

First consider the stress tensor correlations. 
Although the spectral density in the strongly  
coupled theory is markedly different from the free
theory, in Euclidean space-time the correlators differ 
only by 10\%.  
This 10\% difference is significantly 
smaller than the errors associated with this correlator 
on the lattice \cite{nakamura97,nakamura05}.  
At least from the perspective of the AdS/CFT correspondence,
it is hopeless to measure the transport properties 
of the medium in the $T^{\mu\nu}$ channel where 
noisy gluonic operators dominate the signal. 
The figure also suggests that armed only 
with Euclidean measurements 
it is difficult to tell whether the theory
is strongly interacting, {\it i.e.} whether  there is 
a transport peak in the spectral density. 
Therefore, at least within the narrow framework of this
work, rough agreement between quasi-particle
calculations and lattice results \cite{Blaizot:2000fc,Andersen:1999fw} says
 little about the validity of the quasi-particle picture. 
Only  precise agreement can firmly establish validity.

In weak coupling, the insensitivity to the 
transport time scale in the Euclidean correlator 
is readily understood from \Eq{gyxyx_2} and
the free correlator Eqs.~(\ref{freerhotautaua_text})--(\ref{freerhotautauc_text}) \cite{teaneyd,aarts02}.
As perturbative interactions are turned on,
the delta function at $\omega \sim 0 $ is
smeared by the inverse transport time scale. 
However, the integral under 
the peak  remains
the enthalpy times an
average quasi-particle velocity squared, $\llangle v_\p^2/5 \rrangle$. 
This is the physics content of the $f$-sum rule
and is derived in \app{sum_rule_sect}. 
When the smeared delta function  is
substituted into \Eq{gyxyx_2}, the Euclidean
correlator is roughly independent of the width of the peak, {\it i.e.} 
independent of the transport time scale. It seems that
the strong coupling correlators remember this insensitivity to
transport at weak coupling.  

For the current-current correlator the
difference between the free and interacting cases is slightly
larger, $\sim 20\%$. Lattice measurements in the electromagnetic
current-current channel are remarkably precise, $\sim 0.5$ -- $2.0\%$ \cite{Peter_light,Peter_heavy,Nakahara:1999vy}.  This precision stems from the fact 
that the current-current channel correlates quark fields 
while the noisy $T^{\mu\nu}$ channel correlates gluonic fields.
With hard work and precise data, the AdS/CFT correspondence would suggest that
some information about transport can be extracted from Euclidean measurements 
in this channel.
However even in the current-current case,
one needs a definite strategy to isolate the low frequency
contributions.
\\
\\
\noindent {\bf Note added.} 
When this work was near completion, I learned that 
P. Kovtun and A. Starinets had nearly completed a similar
computation \cite{Kovtun}.
\\
\\
\noindent {\bf Acknowledgments.} I thank Peter Arnold,
Raju Venugopalan, Andrei Starinets, and Peter Petreczky for useful discussions.  D.~Teaney was supported by grants from
the U.S. Department of Energy, DE-FG02-88ER40388 and DE-FG03-97ER4014.  

\appendix

\section{The Free Theory}
\label{free_section}
The free $\N=4$ Lagrangian is written as follows:
\begin{equation}
\Ell = 2 \tr \left\{
-\frac{1}{4}F^2 + \frac{1}{2} \bar{\lambda}_a (-i
\bar{\sigma} \cdot \partial )  \lambda^a - \frac{1}{2} \partial_{\mu} X_i \partial^{\mu}X_i  \right\} \, ,
\end{equation}
where ``a" is a SU(4) index and ``i" is a SO(6) index. 
Under flavor rotation,  $\lambda^a$ transforms
in the fundamental representation of SU(4) and 
$X^i$ transforms as the fundamental representation of 
SO(6). SU(4) and SO(6) are locally isomorphic. 
SU(4) matrices are parametrized as $e^{i\beta_A (T^4)_A}$ 
with trace normalization $\tr[(T^4)_A (T^4)_B] = C_4 \delta_{AB}$ and $C_4 =1/2$.
Similarly,  SO(6) matrices are written as $e^{i\beta_A (T^6)_A}$,
with trace normalization $C_6=1$ \cite{Peskin}. The 
normalization convention adopted here has been fixed 
so that the AdS/CFT correspondence holds at the level of two 
point functions at zero temperature \cite{Freedman:1998tz,Chalmers:1998xr}.

Using  the Noether method we compute the conserved R-charge
current
\st
   J^{\mu}_A =  2\, 
\tr\left\{
\left(\frac{T_A}{2}\right)_{ij}  X_i {-i \overleftrightarrow{\partial^{\mu}}} X_j + 
                (T_A)_{ab} \bar{\lambda}_a \left( \frac{\bar{\sigma}^{\mu}}{2} \right) \lambda_b
\right\} \, ,
\stp
where $\overleftrightarrow{\partial} \equiv \overrightarrow{\partial} - \overleftarrow{\partial}$ and gives symmetric Feynman rules.
The spectral density $\rho_{JJ}^L(\omega)$ is easily 
computed at $\k=0$;
the details of a similar computation appear explicitly in an appendix of \Ref{teaneyd}.
The full spectral density is 
\st
     \rho_{\JJ}^{L}(\omega) = 
\left(\rho_{\JJ}^{L}\right)_{\rm fermion}  +
\left(\rho_{\JJ}^{L}\right)_{\rm scalar} \, ,
\label{freerho1}
\stp 
with 
\bg
    (\rho_\JJ^{L})_{\rm scalar} &=& 
\left(
\frac{\chi_s^0}{3}
\right)_{\scriptscriptstyle X} 
\omega\delta(\omega)  + 
                  \frac{N^2 C_6}{96 \pi^2} \, (1 + 2n_B(\omega \beta/2))\,\omega^2  \, , \\
    (\rho_\JJ^{L})_{\rm fermion} &=& 
\left(
\frac{\chi_s^0}{3}
\right)_{\scriptscriptstyle \lambda} 
\omega\delta(\omega)  + 
                  \frac{N^2 C_4}{24 \pi^2}\, (1 - 2n_F(\omega \beta/2))\,\omega^2  \, .
\label{freerho2}
\nd
Here $\left(\chi_s^0\right)_{\scriptscriptstyle X} = \frac{1}{6} \, N^2
T^2 C_6$ and $\left(\chi_s^0\right)_{\scriptscriptstyle \lambda} =
\frac{1}{6} N^2 T^2 C_4$ are the free R-charge static susceptibilities
associated with the scalars and fermions respectively.

Similarly, we compute the spectral density of the stress tensor correlations.
We first construct  the  canonical stress tensor using 
the Noether method  and then construct
the symmetric traceless Bellifante tensor as described in
Weinberg's book \cite{Weinberg}.
The full stress tensor is written
\st
 T^{\mu\nu} = \left(T^{\mu\nu}\right)_{\rm gauge}
                + \left(T^{\mu\nu}\right)_{\rm fermion}
                 + \left(T^{\mu\nu}\right)_{\rm scalar} \, ,
\stp
with
\bg
  \left(T^{\mu\nu}\right)_{\rm gauge}  &=&  2\, \tr \left\{ F^{\mu}_{\kappa} F^{\nu \kappa} 
  + g^{\mu\nu}\left( -\frac{1}{4}F^2 \right) \right\} \, ,
\label{freerhoa} \\
  \left(T^{\mu\nu}\right)_{\rm fermion} &=& 
2\,\tr\left\{ 
\frac{i}{8} 
\bar{\lambda} \left( \bar{\sigma}^{\mu} \overleftrightarrow{\partial^\nu}  + \bar{\sigma}^{\nu} \overleftrightarrow{\partial^\mu} \right) \lambda + 
g^{\mu\nu} \left(\frac{1}{2}
\bar{\lambda}^{a}(-i\bar{\sigma}\cdot \partial) \lambda_a \right) \right\} \, , \\
  \left(T^{\mu\nu}\right)_{\rm scalar} &=& 
2\,\tr\left\{ 
          \partial^{\mu}X_{i} \partial^{\nu} X_{i} +
   g^{\mu\nu} \left(-\frac{1}{2} \partial_{\alpha} X_i \partial^{\alpha} 
X_i \right) 
\right\}
\,.
\label{freerhob}
\nd

Now let us compute $\rho_{\tau\tau}^{yxyx}$.
A straightforward (though lengthy) one loop computation  
calculation leads the spectral density $\rho_{\tau\tau}^{yxyx}$.
The  spectral density is a sum of the scalar, fermion and gauge
boson contributions
\st
     \rho_{\tau\tau}^{yxyx}(\k,\omega) = 
\left(\rho_{\tau\tau}^{yxyx}\right)_{\rm gauge} + 
\left(\rho_{\tau\tau}^{yxyx}\right)_{\rm fermion}  +
\left(\rho_{\tau\tau}^{yxyx}\right)_{\rm scalar} \, .
\stp
These contributions are
\begin{eqnarray}
\left(\rho_{\tau\tau}^{yxyx}\right)_{\rm gauge}
&=& 
\left(\frac{e + p}{5}\right)_{\scriptscriptstyle A} \omega \delta(\omega) +  
\frac{N^2}{160\pi^2} \,\left(1 + 2\,n_B\left(\omega \beta/2 \right)\right) \, \omega^4   \, ,
\label{freerhotautaua} \\
\left(\rho_{\tau\tau}^{yxyx}\right)_{\rm fermion} &=&   
\left(\frac{e + p}{5}\right)_{\scriptscriptstyle \lambda} \omega \delta(\omega) +  
\frac{N^2}{160\pi^2} \,\left(1 - 2\,n_F\left(\omega \beta/2 \right)\right) \, \omega^4  \, , \\
\left(\rho_{\tau\tau}^{yxyx}\right)_{\rm scalar}  &=&   
\left(\frac{e + p}{5}\right)_{\scriptscriptstyle X} \omega \delta(\omega) + 
\frac{N^2}{320\pi^2} \,\left(1 + 2\,n_B\left(\omega \beta/2 \right)\right) \, \omega^4 \, , 
\label{freerhotautauc}
\end{eqnarray}
where for example $(e+p)_{\scriptscriptstyle X}$ is the partial
enthalpy due to the free scalar fields. Explicitly the partial
enthalpies  are 
$(e+p)_{\scriptscriptstyle A} = \frac{4\pi^2}{45} N^2 T^4$ , 
$(e + p)_{\scriptscriptstyle \lambda} = \frac{14 \pi^2}{45}  N^2 T^4$, 
and
$(e + p)_{\scriptscriptstyle  X} = \frac{12 \pi^2}{45} N^2 T^4$.

Eqs.~\ref{freerhoa}--\ref{freerhob} and Eqs.~\ref{freerhotautaua}--\ref{freerhotautauc} constitute the free spectral functions for 
the current-current  and 
the tensor-tensor correlators.  

\section{WKB solution for large $\omega$}
\label{large_omega}
When $\wb$ is large we may solve \Eq{stareq} by a WKB type 
approximation. We first introduce a change of variables
\st
          \psi(u) \equiv \sqrt{ \frac{1-u^2} {u} }\phi(u) \; ,
\label{coordinates}
\stp
which obeys a Scrh\"odinger equation
\st
   \frac{ d^2 \psi}{du^2}  + \frac{1}{u^2(1-u^2)}\,
              \left(-3 + 4 \wb^2 u + 6u^2 + u^4\right) \psi = 0  \; .
\stp
This equation has two singular points at $u=0$ and $u=1$.  
Away from  the singular points we obtain the two WKB solutions,
\bg
   \psi_1  &\sim& \frac{1}{\sqrt{p(u)}}\, \cos( S(u) + \phi_1)\; ,  \\
   \psi_2  &\sim& \frac{1}{\sqrt{p(u)}}\, \sin( S(u) + \phi_2)\; ,
\nd
where $p(u)$ and $S(u)$ are the analogs of momentum  and 
action and  are given by
\bg
    p(u) &=& \frac{ \wb }{\sqrt{u} (1- u^2) } \, ,\\
    S(u) &=& \int_0^{u} p(u')\, du' = \wb \,\tan^{-1}(\sqrt{u}) + \wb\, \tanh^{-1} (\sqrt{u} ) \, .
\nd

Following the WKB strategy we will solve the equation exactly 
near the singular points. Near $u=0$ we expand the potential 
and obtain the following equation 
\st
     \frac{d^2 \psi}{du^2} + \left( -\frac{3}{4u^2} + \frac{\wb^2}{u} \right) \psi = 0  \; .
\stp
The general solution to this equation is 
\st
    \psi(u) =  C_1 \sqrt{u}\, J_2(\sqrt{4\wb^2 u}) +  C_2 \sqrt{u}\, N_2(\sqrt(4\wb^2 u) \; .
\stp
Using the asymptotic expansions of the Bessel functions we 
obtain the matching formulas
\bg
\label{connect1_a}
   \sqrt{u} J_2(2\wb\sqrt{u}) &\rightarrow& \frac{1}{\sqrt{\pi}} 
\frac{1}{\sqrt{p(u)}} \cos \left(S(u) - \frac{5\pi}{4} \right) \; ,\\
   \sqrt{u} N_2(2\wb\sqrt{u}) &\rightarrow& \frac{1}{\sqrt{\pi}} 
\frac{1}{\sqrt{p(u)}} \sin \left(S(u) - \frac{5\pi}{4} \right) \; . 
\label{connect1_b}
\nd
In obtaining these formulas we have used the expansion of 
$p(u)\approx \wb/\sqrt{u}$ and $S(u)\approx 2\wb\sqrt{u}$ near zero. 

Near $u=1$ we solve the equation
\st
\frac{d^2\psi}{du^2} + \frac{1}{4(1-u)^2} \left(1 + \wb^2\right)\psi = 0 \; ,
\stp
and obtain
\st
  \psi = C_1 \, (1 - u)^{\frac{1}{2} - i\frac{\wb}{2}} 
       + C_2 \, (1 - u)^{\frac{1}{2} + i\frac{\wb}{2}} \; .
\stp
The physical solution (transformed according to \Eq{coordinates}) contains only $(1-u)^{\frac{1}{2} - i\frac{\wb}{2}}$.  Now it is a simple matter to show that 
\st
\label{connect2}
\frac{1}{\sqrt{\pi p(u)}}\left[ \cos \left(S(u) - \frac{5\pi}{4} \right) + i \sin \left(S(u) - \frac{5\pi}{4}\right) \right] \stackrel{u\rightarrow1}{\longrightarrow}\mbox{Const}\; (1-u)^{\frac{1}{2} - i\frac{\wb}{2}} \, .
\stp
The constant is irrelevant  to what follows.

Putting together
the connection formulas, Eqs.~(\ref{connect1_a}), (\ref{connect1_b}) and (\ref{connect2}), we find that 
\st
  u J_{2}(2\wb\sqrt{u}) + i u N_2(2\wb\sqrt{u}) 
          \stackrel{\u\rightarrow1}{\longrightarrow} \mbox{Const} \;
                 (1-u)^{- i\frac{\wb}{2}} \, .
\stp
Up to a normalization constant this is the physical solution.
The normalization is fixed by the requirement 
$\left.\phi(u)\right|_{u=0}=1$  and the 
the series expansions, 
$u N_2(2\wb u) \approx \frac{-1}{\pi \wb} + O(u)$ and 
$u J_2(2\wb \sqrt{u)}  \approx  \frac{1}{2}\,\wb^2 u^2  + O(u^3)$\,.
Thus the physical solution in a neighborhood of $u=0$  is
\st
     \phi_1(u) =
 -\pi \wb^2 u N_2(2\wb\sqrt{u})  + i \pi \wb^2  u J_{2}(2\wb\sqrt{u}) \, .
\stp
  Substituting $\phi_1(u)$ into \Eq{GRTxy} we obtain
the leading behavior\footnote{
In \Fig{rho} this result is rewritten as
$\pi \rho(\omega)/\omega = \eta_{\scriptscriptstyle AdS} \,\pi \wb^3 $.
with $\eta_{\scriptscriptstyle AdS} = (\pi N^2 T^3)/8$.}
\st
\label{rho_tau_0T}
      \rho_{\tau\tau}^{yxyx}(\omega) =  \frac{N^2}{64\pi^2} \, \omega^4 \, .
\stp
This is the leading behavior at large frequency 
and agrees with the zero temperature result \cite{Gubser:1998bc}.  The 
WKB solution provides an excellent check of 
the numerical work.

A similar WKB analysis applies to the R-charge correlator;
the details are omitted. 
At large frequency the spectral
density is
\st
\label{rho_j_0T}
   \rho_\JJ(\omega) = \frac{N^2}{32 \pi^2} \omega^2 \, .
\stp


\section {Hydrodynamic Modes} 
\label{modes}

In this appendix we the review the hydrodynamic
modes to keep the treatment self contained. 
Consider a small perturbation from equilibrium.
The stress tensor can be written as the equilibrium
stress tensor plus small corrections 
\begin{eqnarray}
\label{perturb}
   \llangle T^{00}  \rrangle  &=& e + \epsilon(\x,t)\,, \\
   \llangle T^{0i}  \rrangle  &=& 0 + g^i(\x,t) \, .
\end{eqnarray}
The velocity is small and is $\v \equiv \g/(e + p)$.

The linearized hydrodynamic equations are
\begin{eqnarray}
   \partial_t \epsilon + \partial_i g^{i} &=&0 \,,  \\
   \partial_t g^{j} + \partial_i \tau^{ij} &=& 0 \,,
\end{eqnarray}
where  $\tau_{ij}$ is
\begin{eqnarray}
   \tau^{ij} = \delta^{ij} p -  
            \eta (\partial^i v^j + \partial^j v^i  - \frac{2}{3}\delta^{ij} \partial_l v^l )  - \delta^{ij} \zeta \partial_l v^l  \, ,
\end{eqnarray}
and we use metric $(-,+,+,+)$ in this section.
To solve these equations we first take spatial Fourier
transforms, $ \g_\k(t)  = \int d^3\x \,e^{-i\k\cdot\x} \, \g(\x,t)$\,.
Next we divide the momentum vector into transverse and longitudinal
pieces,
\[
     \g_\k  = \gt_\k + \hat{\k}\, \gl_\k \;,
\]
where $\hat{\k} \cdot \gt_\k = 0$. 
For $\gt_\k(t)$, the solution of these linearized equations is
\st
\label{gtr}
     \gt_\k(t) = \gt_\k (0)\, e^{- \frac{\eta k^2}{e +p} \,t } \, ,
\stp
where $\gt_\k(0)$ is the initial condition.
For  $\epsilon_\k(t)$ and $\gl_\k(t)$ the solutions are
\begin{eqnarray}
\label{glong_a}
 \epsilon_\k(t) &=& e^{-\frac{1}{2} \Gamma_s k^2 t} \left[ \epsilon_\k(0) \cos(c_s k t) - i\left(\frac{\gl_\k(0)}{c_s} + i \frac{\Gamma_s k}{2c_s }\epsilon_\k(0)\right)\,\sin(c_s k t)   \right] \, , \\
\label{glong_b}
 \gl_\k(t) &=& e^{-\frac{1}{2} \Gamma_s k^2 t} \left[ \gl_\k(0) \cos(c_s k t) - i\left(c_s \epsilon_\k(0) - i \frac{\Gamma_s k}{2c_s }\gl_\k(0)\right)\,\sin(c_s k t)   \right]  \, .
\end{eqnarray}


To connect these solutions with correlators we
follow the framework of linear response \cite{Forster,teaneyd}. 
The definitions of the correlators used below
are given in \Sect{stress-tensor}. 
We slowly turn
on a small velocity field with a perturbing Hamiltonian
\st
\label{hamiltonian}
   H = H_{0} - \int d^3\x\, v^i(\x,t) T^{0i}(\x,t) \, ,
\stp
and  switch it off at $t=0$. $v^i(\x,t)$ obeys
\st
  v^i(\x,t) = e^{\epsilon t} \theta(-t) v_0^i(\x) \, .
\stp
From the framework of linear response we have 
\st
    \partial_t \llangle g^i(\k, t)\rrangle  = - \,\chi_{gg}^{ij}(\k,t) v_0^i(\k) \,.
\stp
Writing $\v_0(\k) = \v_0^{T}(\k) + \hat{\k} v_0^L(\k)$, and 
substituting the tensor decomposition \Eq{chigg} into this equation, 
we obtain
\begin{eqnarray}
\label{linear_response_1}
    \partial_t \llangle \gl(\k,t) \rrangle &=& - \chi_{gg}^L(\k,t) \, v^L_0(\k)  \, ,\\
\label{linear_response_2}
    \partial_t \llangle \gt(\k,t) \rrangle &=& - \chi_{gg}^T(\k,t) \, \v_0^T(\k) \, .
\end{eqnarray}

The velocity field can be eliminated in favor of the 
the initial values  
\begin{eqnarray}
\label{static_gg}
   \llangle \g^i(\k, 0) \rrangle &=& \chi_{s,gg}^{ij}(\k) v^j_0(\k)  \, , 
\end{eqnarray}
with the static susceptibility 
\st
\chi_{s,gg}^{ij} = \int_0^{\infty} dt\,e^{-\epsilon t} 
\int d^3\x \, e^{-i\k\cdot \x}
\,\llangle \left[ T^{0i}(\x,t), T^{0j}(0,0) \right] \rrangle \, .
\stp
As in \Eq{chigg} the static susceptibility is also broken
up into longitudinal ($\chi_{s,gg}^L$) and transverse 
($\chi_{s,gg}^{T}$) components.

Provided the wavelength of the perturbing velocity field
is long, the system is in perfect equilibrium at time 
$t=0$.
The statistical  
operator describing this system at finite velocity is
\st
   \rho(T,\v_0) = e^{-\frac{(P^0 - v^i_0 P^i)}{T} } \, .
\stp
The stress energy tensor $T^{\mu\nu} = (e + p)\,u^{\mu}u^{\nu} + p \,g^{\mu\nu}$ is
\begin{eqnarray}
\label{stress0}
  \llangle T^{00} \rrangle  = e \qquad \llangle T^{ij} \rrangle  = p \,\delta^{ij} \qquad \llangle T^{0i} \rrangle \equiv g^i = (e + p) v^i_0 \,,
\end{eqnarray}
In this long wavelength limit then, the static susceptibility
is from \Eq{static_gg} and \Eq{stress0}
\st
   \chi_{s,gg}^{ij} = \frac{\partial \llangle g^i \rrangle}{\partial \llangle v^j_0 \rrangle } = (e + p) \, \delta^{ij} \, ,\\
\stp
or $\chi_{s,gg}^L = \chi_{s,gg}^T = e + p$. \,
It is also clear  that the perturbing Hamiltonian does not 
change the average energy. This means that  
$\epsilon(\x,0) =0$ in \Eq{perturb}. 

From this discussion we first, eliminate  $\v$ from the
equations
\begin{eqnarray}
   \partial_t \llangle \gl(\k,t) \rrangle &=& -\chi_{gg}^L(\k,t) 
\frac{ \llangle \gl(\k,0) \rrangle }{e + p } \, ,\\
   \partial_t \llangle \gt(\k,t) \rrangle &=& -\chi_{gg}^T(\k,t) \, 
\frac{ \llangle \gt(\k,0) \rrangle }{e + p } \, .
\end{eqnarray}
Comparing this result with the solution of the hydrodynamics equations  given in \Eq{gtr} and \Eq{glong_b} (with $\epsilon_\k(0) = 0$), 
we deduce the retarded correlators
\begin{eqnarray}
   \chi_{gg}^{L}(\k, t) &=& (e + p)  \, e^{-\frac{1}{2}\Gamma_s k^2 t}\, \left[ c_s k \sin(c_s k t) + \Gamma_s k^2 \,\cos(c_s k t) \right] \,, \\
   \chi_{gg}^{T}(\k, t) &=& \eta k^2\, e^{-\frac{\eta k^2}{e + p} t } \,.
\end{eqnarray}
To find the retarded correlator in frequency 
space, we  integrate $\int_0^{\infty} e^{+i\omega t}$
and find 
\bg
   \chi_{gg}^L (\k, \omega) &=& \frac{i\omega \Gamma_s k^2 - (c_s k)^2}
{\omega^2 - (c_s k)^2 + i\omega \Gamma_s k^2} \, , \\
   \chi_{gg}^T (\k, \omega) &=& \frac{ \eta k^2}{-i\omega + \frac{\eta k^2}{e+p} } \, .
\nd
The spectral density is defined as the  imaginary 
part of the retarded correlator by $\pi$
\bg
   \frac{\rho_{gg}^L (\k, \omega)}{\omega} 
&=& \frac{e +p}{\pi} 
\frac{  \omega^2 \Gamma_s k^2 }
{(\omega^2 - \csq k^2)^2 + (\omega \Gamma_s k^2)^2 } \, , \\ 
&\simeq& \frac{e +p}{2} 
\left[ 
\frac{1}{\pi} \frac{\Gamma_s k^2/2 }{(\omega - c_s k)^2 + (\Gamma_s k^2/2)^2} + 
\frac{1}{\pi} \frac{\Gamma_s k^2/2}{(\omega + c_s k)^2 + (\Gamma_s k^2/2)^2} 
\right] \, ,\\
   \frac{\rho_{gg}^T (\k, \omega)}{\omega} &=& \frac{1}{\pi} \frac{\eta k^2}
{\omega^2 + \left(\frac{\eta k^2}{e+p}\right)^2 } \, .
\nd

These results are one way to express the Kubo formulas. 
They can be expressed slightly differently using current
conservation. Indeed taking imaginary parts of \Eq{current_conserve_1} and \Eq{current_conserve_2}
we can express the Kubo formula in terms of the 
fluxes
\bg
    \lim_{\omega\rightarrow0} \lim_{k\rightarrow0} \frac{\pi \rho_{\tau\tau}^{xx,xx}(\k,\omega) }{\omega} = \frac{4}{3}\eta + \zeta \, , \\
    \lim_{\omega\rightarrow0} \lim_{k\rightarrow0} \frac{\pi \rho_{\tau\tau}^{yx,yx}(\k,\omega) }{\omega} = \eta  \, .
\nd


\section{The $f$ Sum Rule in weakly interacting theories}
\label{sum_rule_sect}

For a weakly interacting theory the 
transport time scale $\tau_{R}$ is much longer than
the inverse temperature. The spectral
density will have a sharp peak at small frequencies $\omega \sim 1/\tau_{R} \ll T$ 
which reflects this separation of time scales.
We may derive a sum rule for this peak 
following the same steps and caveats as
detailed at length in \Ref{teaneyd}. 
Fourier analysis together with \Eq{current_conserve_1} and \Eq{current_conserve_2} and yields
the relations
\bg
\label{sumrule_start_1}
    2 T\,\int_0^{\Lambda } \frac{d\omega}{\omega} \, \rho^{xx,xx}_{\tau\tau}(\k,\omega) &\approx&  \frac{T}{k^2} \left. \partial_t \chi_{gg}^L(\k, t) \right|_{t \sim 1/\Lambda} \, , \\
\label{sumrule_start_2}
      2T\,\int_0^{\Lambda} \frac{d\omega}{\omega} \, \rho^{yx,yx}_{\tau\tau}(\k,\omega) 
&\approx&  \frac{T}{k^2} \left. \partial_t \chi_{gg}^T(\k, t) \right|_{t \sim 1/\Lambda} \, .
\nd
Here $\Lambda$ is a cut-off that is 
much larger than the inverse transport time scale, but 
small compared to the temperature, $1/\tau_{R} \ll \Lambda \ll T$.
The result of this integral does not depend $\Lambda$ to
first order in the scale separation.

Since the time derivatives in Eqs.~(\ref{sumrule_start_1}) and (\ref{sumrule_start_2}) are evaluated at $t~\sim 1/\Lambda$ 
which is short compared to the collision time $\tau_{R}$, the free
streaming Boltzmann equation should be a good description of the
initial equation of motion. In this effective theory 
$1/\Lambda$ may be taken to zero. 

To evaluate these time time derivatives of the retarded 
correlators, 
we consider the same perturbing Hamiltonian and setup 
as described in \app{modes} -- see Eqs.~(\ref{hamiltonian})--(\ref{linear_response_2}). At time $t=0$ the system is 
in perfect equilibrium with temperature $T$ and velocity $v_0(\x)$.
The thermal distribution function  
\st
   f_0(\x, \p) \equiv \frac{1}{e^{ (E_p - v_{0}^j(\x)\p^j)/T } \mp 1} \approx f_p + \frac{1}{T} \, f_p\,(1 \pm f_p)\, \p^j\,v^j_0(\x) \, ,
\stp
with
 $f_p = 1/(e^{E_p/T} \mp 1)$\,. For short 
times the collision-less Boltzmann equation applies, 
\st
   \left[ \ppx{t} + v_\p^{i} \ppx{x^{i}} \right] f(\x,\p,t) = 0 \,.
\stp
Here
$v_\p \equiv \p/E$ and must not be confused with the velocity
field $v_0(\x)$ and its Fourier transform $v_0(\k)$.
The solution to this equation with the specified initial conditions 
is
\st
\label{collisionless}
   f(\x,\p,t) = f_0(\x - v_\p t, \p)  \, .
\stp
Then
the fluctuation in the momentum density is
\st
   \g^i(\x,t)  = \int \dpslash \, \delta f(\x,\p,t) \p^i\, ,
\stp
with  $\delta f(\x,\p,t) = f(\x,\p,t) - f_p$\;. 
Then taking spatial Fourier transforms with $\k$ conjugate to $\x$ 
and substituting the distribution function, \Eq{collisionless}, we have
\st
\label{freeN}
   \g^i(\k,t)  = \left[\frac{1}{T} \int \dpslash e^{-i\k \cdot v_\p t} f_p(1\pm f_p) \p^{i}\p^{j} \right]  v^{j}_0(\k) \, .
\stp  
For small times, we expand the exponential, and find
\st
\label{NNt}
   \g^{i} (\k,t)  =  \left[(e + p) \delta^{ij} - \frac{1}{2}\,t^2\,k^2\,(e + p) \llangle \frac{v^2_\p}{5}\rrangle \left(\delta^{ij} - 
\frac{k^ik^j}{k^2}\right)  - \frac{3}{2} t^2 k^2\, (e + p) \,
\llangle \frac{v^2_\p}{5} \rrangle \frac{k^ik^j}{k^2} \right] v^j_0(\k)\, ,
\stp  
with 
\st
   (e + p) = \frac{1}{ T} \int \dpslash f_p(1\pm f_p)\, \frac{\p^2}{3}  \, ,
\label{chis_boltzmann}
\stp
and
\st
  \llangle \frac{v^2_\p}{5} \rrangle = 
 \frac{1}{T(e + p)} \,\int \dpslash f_p(1\pm f_p)\, \frac{ \p^2}{3} \,\frac{v_\p^2}{5}  \, .
\label{v2by3}
\stp
Thus, from Eqs.~(\ref{linear_response_1}), (\ref{linear_response_2}), 
(\ref{sumrule_start_1}), (\ref{sumrule_start_2}), and (\ref{NNt}),  
we find 
\bg
2 T\,\int_0^{\Lambda } \frac{d\omega}{\omega} \, \rho^{xx,xx}_{\tau\tau}(\k,\omega)
    &\approx&  T\,(e + p) \llangle \frac{3}{5}\,v^2_\p \rrangle \, ,  \\
2 T\,\int_0^{\Lambda} \frac{d\omega}{\omega} \, \rho^{yx,yx}_{\tau\tau}(\k,\omega) 
    &\approx&  T\,(e + p) \llangle \frac{v^2_\p}{5} \rrangle \, .
\nd
%
%

\end{document}